\documentclass[12pt,preprint]{aastex}
\usepackage{natbib}
\newcommand{\ldl}{$\lambda/{\Delta}{\lambda}$}

\newcommand{\kms}{km~s$^{-1}$}

\begin{document}

\title{\sc Kepler Monitoring of an L Dwarf I. The Photometric Period and White Light Flares}

\author{John E.\ Gizis,\altaffilmark{1} 
Adam J.\ Burgasser, \altaffilmark{2}
Edo Berger,\altaffilmark{3}
Peter K.\ G.\ Williams,\altaffilmark{3}
Frederick J.\ Vrba,\altaffilmark{4}
Kelle L.\ Cruz,\altaffilmark{5,6} 
Stanimir Metchev\altaffilmark{7}}

\altaffiltext{1}{Department of Physics and Astronomy, University of Delaware,  Newark, DE 19716, USA}
\altaffiltext{2}{Center for Astrophysics and Space Science, University of California San Diego, La Jolla, CA 92093, USA}
\altaffiltext{3}{Harvard-Smithsonian Center for Astrophysics, 60 Garden Street, Cambridge, MA 02138, USA}
\altaffiltext{4}{US Naval Observatory, Flagstaff Station, 10391 West Naval Observatory Road, Flagstaff, AZ 86001, USA}
\altaffiltext{5}{Department of Physics and Astronomy, Hunter College, City University of New York, 695 Park Avenue, New York, NY 10065, USA}
\altaffiltext{6}{Department of Astrophysics, American Museum of Natural History, Central Park West at 79th Street, New York, NY 10025, USA} 
\altaffiltext{7}{Department of Physics and Astronomy, State University of New York, Stony Brook, NY 11794, USA}

\begin{abstract}
We report on the results of fifteen months of monitoring the nearby field L1 dwarf WISEP J190648.47+401106.8 (W1906+40) with the Kepler mission. Supporting observations with the Karl G.  Jansky Very Large Array and Gemini North telescope reveal that the L dwarf is magnetically active, with quiescent radio and variable H$\alpha$ emission. A
preliminary trigonometric parallax shows that W1906+40 is at a distance of $16.35^{+0.36}_{-0.34}$ pc, and all observations are consistent with W1906+40 being an old disk star just  above the hydrogen-burning limit. The star shows photometric variability with a period of 8.9 hours and an amplitude of 1.5\%, with a consistent phase throughout the year.  We infer a radius of $0.92 \pm 0.07 R_J$ and $\sin i > 0.57$ from the observed period, luminosity ($10^{-3.67 \pm 0.03} L_\odot$), effective temperature ($2300 \pm 75$K) , and $v \sin i$ ($11.2 \pm  2.2$ \kms). The light curve may be modeled with a single large, high latitude dark spot. Unlike many L-type brown dwarfs, there is no evidence of other variations at the $\gtrsim 2\%$ level, either non-periodic or transient periodic, that mask the underlying rotation period. We suggest that the long-lived surface features may be due to starspots, but the possibility of cloud variations cannot be ruled out without further multi-wavelength observations. During the Gemini spectroscopy, we observed the most powerful flare ever seen on an L dwarf, with an estimated energy of  $\sim 1.6 \times 10^{32}$ ergs in white light emission. Using the Kepler data, we identify similar flares and estimate that white light flares with optical/ultraviolet energies of $10^{31}$ ergs or more occur on W1906+40 as often as 1-2 times per month. 
\end{abstract}


\section{Introduction\label{intro}}

Dramatic changes in the spectra of ultracool dwarfs with $T_{eff} \lesssim 2300$ K lead to their
classification as  a distinct spectral type, the L dwarfs \citep{1999ApJ...519..802K,1999AJ....118.2466M}.  The L dwarf field population is a mix of old hydrogen-burning stars and 
young brown dwarfs. The weakening of molecular features and reddening of broad-band colors compared to M dwarfs are explained by the formation of  condensates, or dust 
grains (see the review by \citealp{2005ARA&A..43..195K}). Early observations established that chromospheric activity weakens in this spectral type range \citep{2000AJ....120.1085G}, and it is
now clear that magnetic activity changes dramatically in character (see \citealp{2010ApJ...709..332B,McLean:2012qy}). L0-L2 dwarfs are cooler than the M dwarfs which show optical variations due to cool magnetic starspots \citep{2011AJ....141...20B}, but warmer than the L/T transition brown dwarfs that show mounting evidence of infrared variations due to cloud inhomogeneties \citep{Artigau:2009ly,Radigan:2012vn,Buenzli:2012lr}. These considerations suggest that L dwarfs may show variability due to 
changes in the condensate distribution (such as clouds, or holes in cloud decks) and/or magnetic starspots.  Because measurements of $v \sin i$ suggest all 
L dwarfs are rapid rotators \citep{Bailer-Jones:2004kx,2008ApJ...684.1390R}, rotational modulation is expected to have periods of hours.   

Optical ($I$-band) variations have been detected in many L dwarfs  at the few percent level 
\citep{2001A&A...367..218B,2002MNRAS.335.1158C,Gelino:2002uq,2003MNRAS.341..239C,2003MNRAS.346..473K,2005MNRAS.360.1132K,2006MNRAS.367.1735K,Lane:2007yq} 
and usually attributed to inhomogenities in the clouds. Although a few detected $I$-band signals were periodic and consistent with rotational modulation, others were not, and characteristics seen in one observing run may be different in another. 
\citet{2001A&A...367..218B} suggested that surface features in early L dwarfs evolve on timescales of hours and ``mask" the underlying rotation curve in $I$-band: A particularly interesting case was \object{2MASSW J1145572+231730} (hereafter 2M1145+23), an L1.5 dwarf with H$\alpha$ emission \citep{1999ApJ...519..802K}, which showed periods that did not match the previous year's observations \citep{1999A&A...348..800B,2001A&A...367..218B}.  \citet{Gelino:2002uq} found that at least seven, and perhaps up to twelve, of eighteen L dwarfs were variable at $I$-band, but that most had non-periodic light-curves. The two observed periods were thought to be much longer than the rotation rates. \citet{2006MNRAS.367.1735K} reported that the L0 dwarf  \object{2MASS J06050196-22342270} had a consistent 2.4 hour period over three nights, but the amplitude declined from 27 to 11 mmag.  Both \citet{2001A&A...367..218B} and \citet{Gelino:2002uq} considered the balance of evidence favored cloud variations over sunspot-like starspots, although some possible starspots have been noted in L dwarfs with H$\alpha$ emission \citep{1999A&A...348..800B,2003MNRAS.341..239C}.  
\citet{Lane:2007yq} suggests that a three-hour periodicity in the L3.5 \object{2MASSW J0036159+182110} is due to starspots: Although this source is not chromospherically active, it is radio active \citep{2005ApJ...627..960B}.  
These observations are all of field L dwarfs; \citet{Caballero:2004uq} did not find statistically significant variations in young ($\sigma$~ Ori) early L-type brown dwarfs, but few percent variations would have been below the detection threshold and warmer young brown dwarfs showed evidence of accretion.   

Flares are another potential type of variability in ultra cool dwarfs, but only a few optical flares in L dwarfs have been reported \citep{2003AJ....125..343L,2007AJ....133.2258S,2008ApJ...684.1390R}. As \citet{2007AJ....133.2258S} remark, only the H$\alpha$ emission line has been seen in these flares, and not the nearby helium emission lines seen in M dwarf flares, though this may be
a matter of sensitivity. 
In a study of M8-L3 dwarfs, \citet{2010ApJ...709..332B} placed a limit of $\lesssim 0.04$ hr$^{-1}$ on the rate of flares that increase the H$\alpha$ line more than a factor of a few above the quiescent chromospheric emission level.  This rate is consistent with the estimated L dwarf flare duty cycle of $\sim 1-2\%$ \citep{2003AJ....125..343L,2007AJ....133.2258S}. More strong flares have been observed in M7-M9 dwarfs \citep{1999ApJ...519..345L,Martin:2001qy,Rockenfeller:2006lr,2006A&A...460L..35S,2007AJ....133.2258S}, with other atomic emission lines and even white light continuum emission in some cases. Their H$\alpha$ flare duty cycle is also much higher, perhaps 5-7\% \citep{2000AJ....120.1085G,2007AJ....133.2258S}. Notably, simultaneous radio-optical monitoring of two M8.5 dwarfs found no temporal correlation between radio flares and H$\alpha$ flares \citep{2008ApJ...676.1307B,2008ApJ...673.1080B}.  

The field L dwarf observations are intriguing, but the limitations of telescope scheduling and ground-based observing make it difficult to reliably characterize L dwarf variability and assess the relative contribution of periodic and non-periodic components. The discovery of \object{WISEP J190648.47+401106.8} (hereafter W1906+40, \citealt{Gizis:2011lr}), a nearby L1 dwarf in the Kepler Mission \citep{2010ApJ...713L..79K} field of view, has allowed us to obtain a 15 month long time series from space. Although W1906+40 is much fainter (SDSS $g,r,i = 22.4, 20.0, 17.4$)  than the typical FGKM main sequence star targeted with Kepler, the achieved precision of 7 mmags per observation compares favorably with ground-based data. We report on the Kepler photometry and supporting multi-wavelength observations in Section~\ref{da}, discuss periodicity in Section~\ref{sec-period}, and discuss flares in Section~\ref{sec-flare}.  
   
\section{Observations\label{da}}

\subsection{Kepler Data}

W1906+40 was observed with Kepler Director's Discretionary Time (Program GO30101) from 28 June 2011 to 27 June 2012 and with Guest Observer time (Program GO40004) from 28 June 2012 to 03 October 2012. Its parameters are listed in Table~\ref{tab1}. It was assigned a new name for each quarter: KIC 100003560 (Quarter 10), KIC 100003605 (Quarter 11), KIC 100003905 (Quarter 12), KIC 100004035 (Quarter 13), and KIC 100004076 (Quarter 14). The target was observed in long cadence mode \citep{Jenkins:2010fk}, providing 30 minute observations, for the full time period and also in one-minute short cadence mode \citep{2010ApJ...713L.160G} during Quarter 14. For each observation, the Kepler pipeline \citep{2010ApJ...713L..87J} provides the pixel data, the time ($t=BJD - 2454833.0$, or mission days), the aperture flux (SAP\_Flux), the uncertainty, and a corrected aperture flux (PDCSAP\_Flux), which is meant to correct instrumental drifts \citep{Stumpe:2012uq}. After excluding data flagged as bad (Quality Flag $>0$), there are 18,372 long cadence measurements.  The pipeline-estimated noise levels are 0.5\%.   The spacecraft is rotated between quarters and the focus is a compromise, so W1906+40 was observed on four different detectors with considerably different point spread functions. The Kepler pipeline therefore selects different apertures for each quarter, which in some cases we can improve upon. For Quarters 10 and 14, we adopt the PDCSAP\_Flux photometry, which used an aperture of nine pixels, because we found no significant improvement with smaller apertures. However, the default pipeline analysis used larger apertures for Quarters 11-13, which resulted in noisier photometry than Quarters 10 and 14. Furthermore, Quarter 13's  PDCSAP\_Flux is very low ($<100$ counts per second), which we found is due to two negative (bad) pixels included in the aperture. Smaller apertures also reduce the possibility of contamination by nearby background stars. We therefore adopted smaller apertures of six pixels for Quarters 11 and 12 and four pixels for Quarter 13 using the PyKE tool kepextract.\footnote{PyKE is available through the Kepler Mission contributed software webpage: \url{http://keplergo.arc.nasa.gov/PyKE.shtml}}  Instrumental drifts for these three quarters were removed with third-order polynomial fits.  Kepler photometry is not calibrated in any absolute sense since such calibration is not needed to meet the primary mission goals, so we divided each quarter's data by the median count rate for that quarter.  

\begin{figure}
\epsscale{0.7}
\plotone{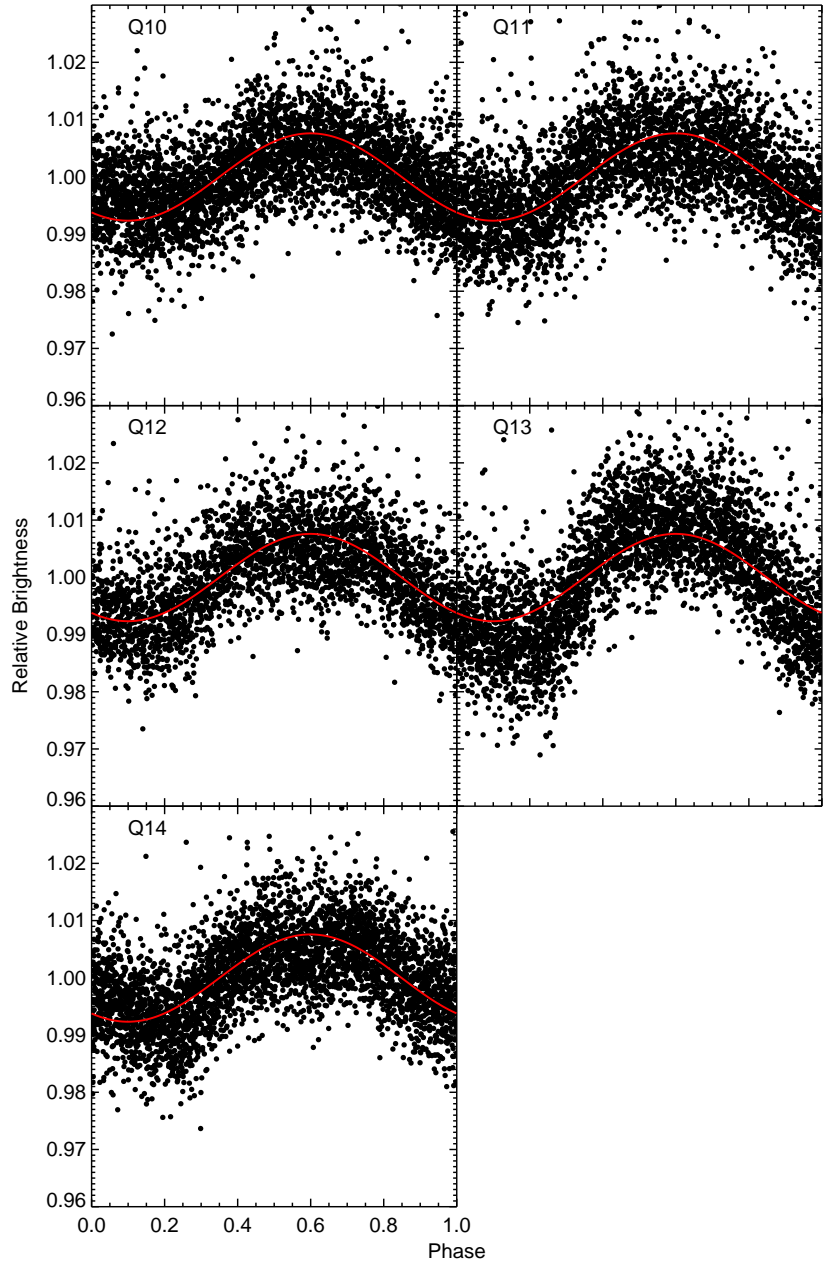}
\caption{Phased Kepler light curves for W1906+40 for Quarters 10-14 for a period of 0.370152 days. Phase zero is defined as Kepler Mission day 1000.0.  The red curve shows a sine function with $P=0.370152$ days and amplitude 0.00763. The light curve shows significant differences in Quarter 13, but overall the light curve has a consistent phase and amplitude.  
\label{fig-phased}}
\end{figure}

\begin{figure}
\plotone{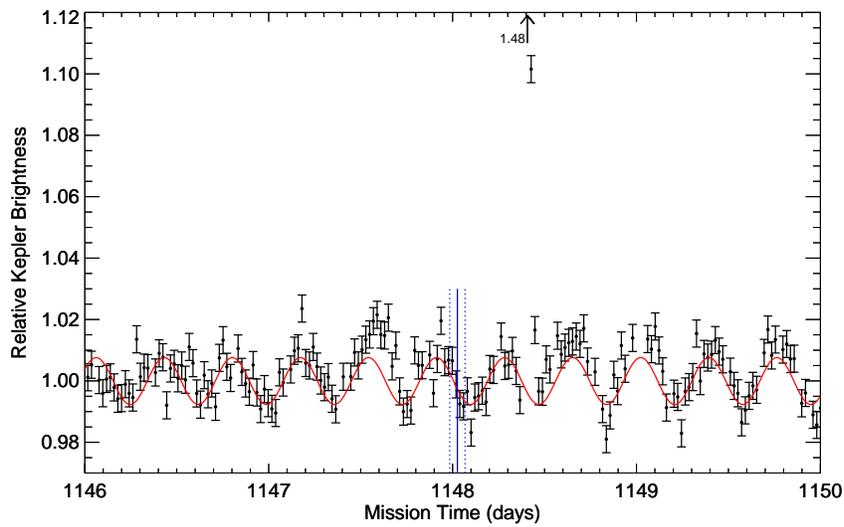}
\caption{Kepler light curve showing the Quarter 12 data from Mission Day 1146 to 1150. The plotted error bars are those reported by the Kepler pipeline (0.044 for these points), which may be underestimated.  The red curve shows the model sine fit for the full year-long dataset. Possible changes in the amplitude are evident, perhaps up to twice the model. The time of the radio observations (Section~\ref{sec-radio}) is marked by the vertical blue line with one hour before and after shown as the dotted line. The flare candidate at 1148.41 is discussed in Section~\ref{sec-kepflare}.
\label{fig-radioregion}}
\end{figure}

The Kepler data show a strong peak in the periodogram at a period of 0.3702 days. We emphasize that this period is present in each quarter, and in both the pipeline photometry and our revised (smaller aperture) photometry. This 8.9 hour period is consistent with the rotation period distribution deduced for early L dwarfs by their $v \sin i$ measurements \citep{2008ApJ...684.1390R}. We fit a sine curve to all five quarters of long cadence data with a Markov Chain Monte Carlo (MCMC) analysis and find the period is $0.370152 \pm 0.000002$ days with an amplitude of $0.00763 \pm 0.00006$ (8.3 mmag). The peak-to-peak variation is twice this (1.5\%, or 16 mmag). Fitting each quarter separately, then averaging, yields a period of $0.37015$ days with a standard deviation of $0.00005$ days. The full phased dataset with sine curve is shown in Figure~\ref{fig-phased}.  Excluding the small number of outliers, whose nature are discussed in Section~\ref{sec-flare}, the standard deviation about the fit is 0.65\%, or 7 mmag. We adopt this as the noise level of the measurements, as it is only a slight increase over the 
expected value (0.5\%) and there are other possible noise sources in the Kepler instrument \citep{2010ApJ...713L..92C}, though real intrinsic variations may contribute. Figure~\ref{fig-radioregion} shows four days of Kepler data as an example of unphased data including a possible flare (Section~\ref{sec-kepflare}); our radio observations (Section~\ref{sec-radio}) were in this time period. The Kepler filter response curve is shown in Figure~\ref{fig-filter} along with the estimated relative count rate at each wavelength due to the L1 dwarf photosphere; possible contributions from flares are discussed in Section~\ref{sec-kepflare}.  

\subsection{Optical and Near-Infrared Observations\label{optical}}

W1906+40 was observed with the Gemini Multi-Object Spectrograph \citep{Hook:2004lr} on the Gemini North telescope (Program GN-2012A-Q-37). The R831 grating, OG515 order blocking filter, and 1.0 arcsecond slit were used to yield wavelength coverage 6276\AA~to 8393 \AA, with two small gaps between detectors that were interpolated over.  We obtained sixteen 600-second exposures on UT Date 24 July 2012, twenty-five on 29 July 2012, and six on 8 August 2012. The flux calibration star BD+28 4211 was observed on 29 July 2012. The data were bias-subtracted, flat-fielded, wavelength-calibrated, extracted and flux-calibrated in the standard way using the Gemini gmos package in IRAF.  A small correction was then applied to bring the spectra in agreement with the $i$-band magnitude.  

\begin{figure}
\plotone{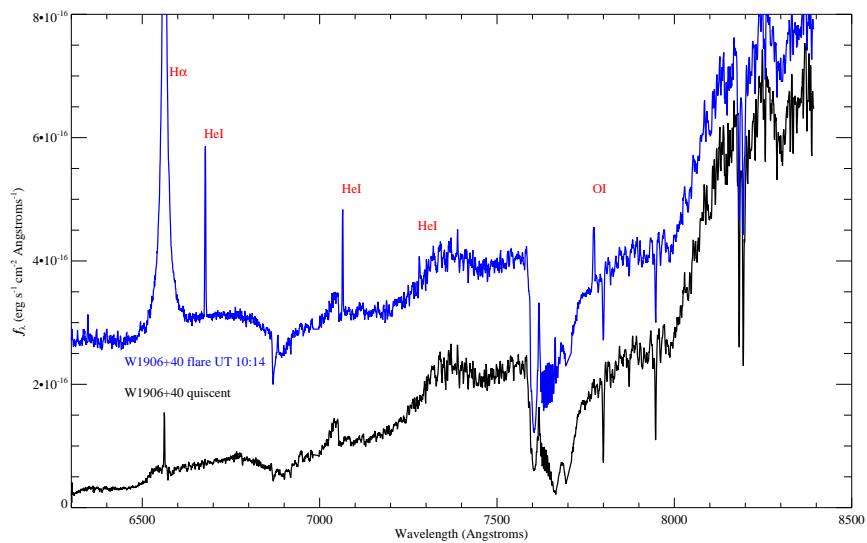}
\caption{Gemini spectrum of W1906+40 (without telluric correction) in a quiescent, apparently non-flaring, period and in a white light flare. The spectral type is L1 on the \citet{1999ApJ...519..802K} system.  
\label{fig-spectrum1}}
\end{figure}

We compared the Gemini spectra to the spectral type standards defined by \citet{1999ApJ...519..802K} and find that W1906+40 is spectral type L1 (Figure~\ref{fig-spectrum1}.) It is not a low gravity L dwarf \citep{2009AJ....137.3345C}.  We see no evidence for lithium absorption in any of the spectra.   The L1 optical spectral type agrees with the L1 near-infrared spectral type derived in the discovery report.  H$\alpha$ is variable on all nights but detected in emission in all spectra. On 24 July the H$\alpha$ emission equivalent width increased from 0.5\AA~ at the beginning of observations to 7.5\AA~ at the end.  On 29 July 2012, the H$\alpha$ emission begins at 2.8\AA, but is then dominated by flares, discussed further in Section~\ref{geminiflare}, as seen in Figure~\ref{fig-spectrum1}. On 8 August, the first five spectra have an equivalent width of 2.5-3.0 \AA, but the last increases to 6.5\AA. The average non-flare H$\alpha$ equivalent width is 4\AA. The flares on 29 July overwhelm the small periodic signal, so we defer any attempt to identify small photospheric changes on the other nights to a future paper when additional simultaneous ground-based and Kepler data are available.

W1906+40 was observed in clear and dry conditions on 10 September 2011 (UT) with the Keck II NIRSPEC near-infrared echelle spectrograph (McLean et al. 2000), as part of an ongoing search for radial velocity variables among nearby  L dwarfs \citep{Burgasser:2012qy}.   The source was observed using the high-dispersion mode, N7 filter and 0$\farcs$432$\times$12$\arcsec$ slit to obtain 2.00--2.39~$\mu$m spectra over orders 32--38 with {\ldl} = 20,000 ($\Delta{v}$ = 15~{\kms}) and dispersion of 0.315~{\AA}~pixel$^{-1}$.  Two 120~s exposures were obtained in two nods separated by 7$\arcsec$ along the slit.  We also observed the A0~V calibrator HD 192538 ($V$ = 6.47) in the same setup for telluric correction and flux calibration, and internal quartz flatfield and NeArXeKr arc lamps for pixel response and wavelength calibration, respectively.

Spectral data were extracted from the images using the REDSPEC package\footnote{REDSPEC was developed for NIRSPEC by S.\ Kim, L.\ Prato, and I.\ McLean; see \url{http://www2.keck.hawaii.edu/inst/nirspec/redspec/index.html}.}, and we focused our analysis on order 33 which samples the strong CO band around 2.3~$\micron$ \citep{Blake:2008qy}. We used REDSPEC and associated routines to perform pixel-response calibration, background subtraction, order identification, image rectification, and extraction of raw counts for both target and A0~V star.  We also obtained an initial linear wavelength calibration using the arc lamp images, but performed no telluric or flux calibration.  We then used these data to perform a forward-modeling analysis similar to that described in \citet{Blake:2010gf}.  Our full methodology will be presented in a future paper, but in brief we explored a parameterized model of the raw spectrum 
\begin{equation}
S[p] = \left\{(M[p(\lambda,RV)]{\star}K){\times}T[p(\lambda)]^{\alpha}{\times}C[p]\right\}{\star}LSF
\end{equation}
through Markov Chain Monte Carlo (MCMC) methods.
Here, M is a synthetic model chosen from the BT-Settled set of \citet{Allard:2011uq},
spanning 1800~K $\leq$ T$_{eff}$ $\leq$ 2500~K in 100~K steps and 4.5 $\leq$ log~g $\leq$ 5.5 (cgs) in 0.5~dex steps, and sampled at a resolution of 300,000 ($\Delta{v}$ = 1 \kms);
$p(\lambda)$ is the conversion function from wavelength to pixel, modeled as a third-order polynomial; 
RV is the barycentric radial velocity of the star; 
K is the rotational broadening kernel defined by \citet{Gray:1992fk} which depends on $v\sin{i}$;
T[$\lambda$] is the telluric absorption spectrum, adopted from the solar absorption spectral atlas of \citet{Livingston:1991fj};
$\alpha$ scales T[$\lambda$] to match the observed telluric absorption;
$C$ is the flux calibration function, modeled as a second-order polynomial;  
LSF is the spectrograph line spread function, modeled as a Gaussian with width $\sigma_G$;
and $\star$ denotes convolution.
To facilitate rapid exploration of this 11-component parameter space, we performed the rotation convolution on all of the models in advance, sampling 5~\kms $\leq$ $v\sin{i}$ $\leq$ 100~\kms in 1~\kms steps.
Parameters were initiated by first fitting the A0~V observation; we then fit the W1906+40 spectrum using MCMC chains of length 1000 for each of the spectral models deployed (T$_{eff}$ and log~g), following standard procedures (e.g., \citealt{Ford:2005qy}) and using $\chi^2$ as our cost function .  

\begin{figure}
\plotone{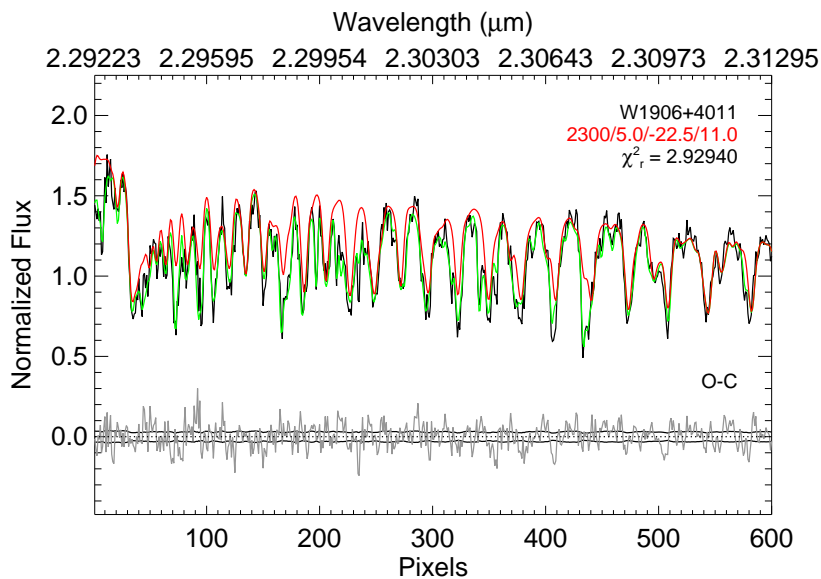}
\caption{Keck II NIRSPEC spectrum of W1906+40 with best-fit model to the data.
\label{fig-bestspec}}
\end{figure}

Figure~\ref{fig-bestspec} shows our best-fit model to the data, with T$_{eff}$ = 2300~K, log~g = 5.0~dex, heliocentric radial velocity of $-$22.5~\kms (assuming a barycentric motion of $-$10.7~\kms) and $v\sin{i}$ = 11~\kms.  The fit is good (reduced $\chi^2$ = 2.9), and the strong blend of both stellar and telluric absorption features provides a robust framework for constraining both radial and rotational velocities.  To determine estimates of each, we marginalized all of the chains for each parameter $p$ using a probability function based on the F-test statistic
\begin{equation}
\langle{p}\rangle = \sum_i^{\rm all~chains}{p_i\wp_i}
\end{equation}
\begin{equation}
\sigma^2_p = \left(\sum_i^{\rm all~chains}{p_i^2\wp_i}\right) - \langle{p}\rangle^2 
\end{equation}
where the probability function $\wp$ for each chain was determined from the F-test probability distribution function
\begin{equation}
\wp_i \propto 1-F(\chi^2_i/{\rm min}(\{\chi^2\}),\nu,\nu)
\end{equation}
(see \citealt{Burgasser:2010fk}) with ${\rm min}(\{\chi^2\})$ being the minimum $\chi^2$ (best-fitting model), $\nu$ the degrees of freedom (\# pixels $-$ \# parameters), and enforcing $\sum\wp_i = 1$.  From this calculation we determined RV = $-$22.6$\pm$0.4~\kms~ and $v\sin{i}$ = 11.2$\pm$2.2~\kms~ for W1906+40. The fitted $T_{eff}$ is consistent with other L dwarf studies (see the reviews of \citealt{2000ARA&A..38..337C,2005ARA&A..43..195K}), so we estimate the uncertainty in $T_{eff}$ as $\pm 75$K.

The parallax and proper motion were obtained with the ASTROCAM \citep{Fischer:2003fj} imager at the 1.55-m telescope of the U.S. Naval Observatory, using the astrometric observing and reduction principles described in  \citet{2004AJ....127.2948V}. Observations in J-band were obtained on 27 nights over the course of 1.23 years. A reference frame employing 16 stars with a range of apparent J magnitude of 11.8-14.6 was employed. The conversion from relative to absolute parallax was based on using both 2MASS and SDSS photometry to determine a mean distance of the reference frame stars of 2.20 mas. We caution that this is a preliminary parallax since, even with an excellent reference frame and a location in the sky giving a significant parallax factor in declination, an observational time baseline of 1.23 years is not sufficient to completely separate parallax and proper motion. This object will be on the USNO infrared parallax program for the next 2-3 years to establish a final parallax.  This distance ($16.35^{+0.36}_{-0.34}$ pc) is in excellent agreement with that derived in the discovery paper ($16.6 \pm 1.9$ pc) assuming the object to be single. If there is a unresolved companion, it is therefore unlikely  to contribute significantly to the Kepler photometry. As shown in the \citet{2005ARA&A..43..195K} review, an L1 dwarf like W1906+40 is most likely a hydrogen-burning star with age $\gtrsim 300$ Myr. The observed tangential velocity is greater than the median (30 \kms) for L1 dwarfs \citep{2009AJ....137....1F}, supporting an old age and stellar status. 
Both integrating the W1906+40 near-infrared spectrum and optical to mid-infrared photometry and comparison to 
other L1 dwarfs \citep{2004AJ....127.3516G} gives $BC_K =3.22 \pm 0.06$, and therefore $\log L/L_\odot = -3.67 \pm 0.03$. Overall, our optical and near-infrared observations show that W1906+40 is a typical L1 dwarf, though the H$\alpha$ and rotation rate places it in the more chromospherically active and slowly rotating half of the population.  If the L dwarf rotation-age relation suggested by \citet{2008ApJ...684.1390R} is correct, W1906+40 would be older than 5 Gyr.

\subsection{Radio Observations\label{sec-radio}}

W1906+40 was observed with the Karl G. Jansky Very Large Array (VLA)\footnote{The VLA is operated by the National Radio Astronomy
Observatory, a facility of the National Science Foundation operated under cooperative agreement by Associated Universities, Inc.} 
for three hours starting at 2012 Feb 23.46~UTC (Project 12A-088). The total observing bandwidth was 2048~MHz, with two subbands centered at 5000 and 7100~MHz and divided
into 512 channels each. Standard calibration observations were obtained, with 3C\,286 serving as the flux density scale and bandpass calibrator
and the quasar J1845+4007 (4\degr\ from W1906+40) serving as the phase reference source. The total integration time on W1906+40 was 120~minutes.

The VLA data were reduced using standard procedures in the CASA software system \citep{2007ASPC..376..127M}. Radiofrequency interference was
flagged manually. After calibration, a deep image of 2048$\times$2048 pixels, each 1$\times$1 arcsec, was created. The imaging used
multi-frequency synthesis \citep{1994A&AS..108..585S} and CASA's multifrequency CLEAN algorithm with two spectral Taylor series terms for each CLEAN component; this approach models both the flux and spectral index of each source. The rms residual of the deconvolution process around W1906+40 was 2.9~$\mu$Jy~bm$^{-1}$. 

W1906+40 is detected at the eight-sigma level with flux density $23.0 \pm 4.1$ $\mu$Jy at a mean observing frequency of 6.05 GHz. 
This corresponds to $\nu L_\nu = (4.5  \pm 0.9) \times 10^{22}$ erg s$^{-1}$ with spectral index 
($S_\nu \propto \nu^\alpha$) $\alpha =  -2.0 \pm 1.4$.  The mean time of the observation is
23 Feb 2012 12:40:50 UTC which equals Kepler mission time 1148.026. To place the radio coverage in context, the time of the radio observations is marked on the corresponding Kepler photometry in Figure~\ref{fig-radioregion}; the mean time of the radio observations corresponds to phase 0.91 in Figure~\ref{fig-phased}, close to the minimum in the Kepler light curve. We searched for radio variability in W1906+40 by using the deep CLEAN component model to subtract all other detectable sources from the visibility data, then plotting the real component of the mean residual visibility as a function of time, rephasing the data to the location of W1906+40. No evidence for radio variability was seen. We imaged the LL and RR polarization components separately and
found no evidence for significant circular polarization of the emission from W1906+40. W1906+40 is just the sixth L/T dwarf detected in quiescent radio emission \citep{Berger:2002fk,2006ApJ...648..629B,2009ApJ...695..310B,Burgasser:2013qy,Williams:2013fk}.  

\subsection{Mid-Infrared (WISE) data}

The W1906+40 discovery paper was based on the Preliminary Wide Field Infrared Explorer (WISE, \citealp{2010AJ....140.1868W}) data release. It appears as source WISE J190648.47+401106.8 in the final WISE All-Sky Source Catalog. The source was scanned 28 times over 1.9 days in April 2010 as part of the main WISE Mission and again 34 times over 4.3 days in October 2010 in the Post-Cryo mission. We find that in the WISE All-Sky Single Exposure (L1b) Source Table and WISE Preliminary Post-Cryo Single Exposure (L1b) Source Tables, W1 is consistent between and within the two epochs, but the source apparently brightened from W2$=11.23 \pm 0.02$ in April 2010 to a mean W2$=11.12 \pm 0.02$ in October (with individual uncertainties $\pm 0.03$ magnitudes.) The October data also shows more scatter in W2 than expected, but higher precision mid-infrared observations would be needed to confirm variability. The WISE data are more than 680 periods before the beginning of the Kepler data.

\section{Photometric Variability \label{sec-period}}

\subsection{Radius and Inclination}

The radius and inclination of W1906+40 are important parameters in interpreting the Kepler light curve. We can constrain them both using our observations.  
First, the radius is a function of the observed distance (and thereby the luminosity) and our effective temperature fit:

\begin{equation}
\label{eqn-radius2}
R = 0.90 \pm 0.03 R_J \left(\frac{2300 {\rm K}}{T_{eff}}\right)^2 \left(\frac{d}{16.35 {\rm pc}}\right) 
\end{equation}

The uncertainty of 3\% includes the uncertainty in the bolometric correction and photometry but not parallax or temperature uncertainty. Second, assuming the photometric period is the rotation period, we can also constrain the radius and inclination with our spectroscopic $v \sin i$: 

\begin{equation}
\label{eqn-radius1}
R \sin i= 0.80 R_J \left({P}\over{0.37015 ~\rm{d}}\right)
\left({v \sin i}\over{11.2 ~\rm{km s^{-1}}}\right) 
\end{equation} 

\begin{figure}
\plotone{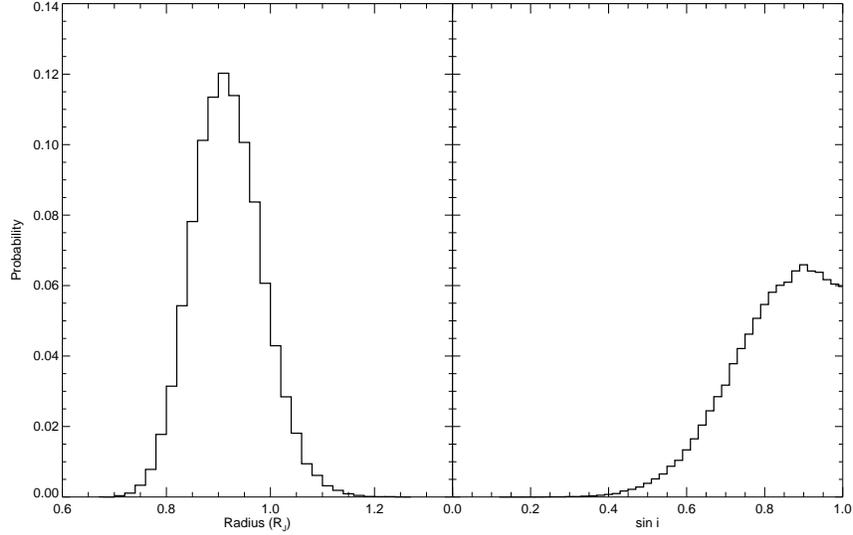}
\caption{Posterior Probability Distribution for the radius and inclination ($\sin i$) of W1906+40 given the constraints of the observed luminosity, effective temperature, period, and $v \sin i$.  
\label{fig-Lprob}}
\end{figure}

The uncertainty in this value is 20\% due to the $v \sin i$ uncertainty. Combining this constraint with Equation~\ref{eqn-radius2} suggests $i \approx 60^\circ$, but a range of inclinations are possible.    
Since the period is so strongly constrained, we can infer the posterior probability using three observations for the data (luminosity, effective temperature, and $v \sin i$) and a three parameter model (luminosity, radius, and $\sin i$). Our priors are flat in luminosity and radius, but proportional to $\sin i$ on geometric grounds. Using an MCMC chain of one million steps, and discarding a burn-in of one thousand steps, we find the posterior probability distribution shown in Figure~\ref{fig-Lprob}. The inferred radius is $R = 0.92 \pm 0.07 R_J$. At the 95\% confidence level, we find that $\sin i > 0.59$, with the median value $\sin i = 0.83$ and most likely value $\sin i = 0.90$.  The inferred radius is consistent with theoretical predictions of $0.8 \lesssim R \lesssim  1.0 R_J$ \citep{2000ApJ...542..464C,Burrows:2011lr}. 

\subsection{Spot models}

The most striking aspect of the Kepler light curve is that it has a consistent phase for over a year.  Furthermore, while we cannot rule out deviations from a sine function or small changes in the shape, it appears that the light curve has no flat portion. The simplest explanation is that the rotating dwarf has a single feature (darker or brighter than the surrounding photosphere) that is always in view. To explore this possibility, we model the first year of data using the circular spot equations of  \citet{Dorren:1987fk}, which we implemented in Python with an MCMC code. We assumed linear limb darkening coefficients of 0.9 based on the predictions of \citet{Claret:1998uq} and \citet{Claret:2011rt}.  Many solutions are possible, and generally we can find single dark spot models with $\chi^2$ similar to the sine models. Some possible solutions with completely dark spots with $\sin i$ consistent with Figure~\ref{fig-Lprob} and  are listed in Table~\ref{tab-spot}.  Solutions with $i \approx 80^\circ$ can be found, but they require a larger spot. The different light curve during Quarter 13 could be explained in large part in the single circular dark spot model by increasing the spot size (to $\sim 13^\circ$ for $i=60^\circ$), or by lowering the latitude (to $\sim 70^\circ$ for $i=60^\circ$). While attributing a long-lived feature to a single ``spot" is attractive, it could also be modeled in other ways, such as multiple large dark spots at lower latitudes.  An intriguing alternative spot model that does not favor large circumpolar spots was presented for active dwarfs by \citet{Alekseev:1996qy}.  In this ``zonal model," it is assumed that a large number of small spots together form large dark bands symmetric around the equator. This naturally produces sinusoidal light curve even at $\sin i \approx 1$ without requiring us to be viewing polar spots.  

The physical cause of the putative large spot at high latitude (or even multiple spots at low latitudes) might be a magnetic starspot (i.e., a region of cooler photosphere due to magnetic fields) or an cloud structure.  Jupiter's Great Red Spot would produce a strong periodic signal at 9.9 hours even in unresolved data \citep{Gelino:2000lr}, though a single spot on W1906+40 would be at much higher latitude. \citet{Gelino:2002uq} comment that small holes in a cloud layer would form bright ($I$-band) spots ``similar to the `5 $\mu$m hot spots' of Jupiter." These might be distributed according to the zonal model. \citet{Harding:2011fk} argue for a ``magnetically-driven auroral process" to produce optical variations in some ultra cool dwarfs. We cannot determine whether the features are brighter or darker than the photosphere with the single-band Kepler data, so all of these possibilities are plausible. It is worth considering whether chromospheric emission line variations would explain the Kepler periods. Varying H$\alpha$ emission from zero to ten Angstroms equivalent width, even with associated bluer Balmer lines, would only produce a signal of $\sim0.0015$\%, an order of magnitude too small to explain our data. 

The stability of the period and amplitude makes a striking contrast with the L dwarf observations discussed in the Introduction. Besides the stable 8.9 hour period, we find no evidence of other periods, transient or not, in the data. To search for other periods of hours to days, we have calculated periodograms for every three, five, and ten day period and found no other significant period in any of them. The long-term drifts we removed were typically 1\%, and only reached 6\% during quarter 11, and these are consistent with the drifts in bright stars corrected by the PDC pipeline process. Although intrinsic slow changes in W1906+40 on timescales of months might be removed, it is likely that a strong signal, like the $\sim30$ day period M dwarf \object{GJ 4099} (KIC 414293, \citealp{2011AJ....141...20B}), would still be detected.  Variations on timescales less than a week that are $\gtrsim 1\%$ would be detected. We conclude that the various ``masking" periods and non-periodic variations reported in other L dwarfs at amplitudes $>2\%$, and attributed to condensate cloud inhomogeneties, are not present in W1906+40. On the other hand, variations $\lesssim 1\%$ can be plausibly be attributed to noise, but could also be due to real variations. (The data shown in Figure~\ref{fig-radioregion} are suggestive, and typical.) 

If the short-lived variability observed on non-active ultracool dwarfs is associated with condensate clouds, perhaps the long-lived periodic feature on this active L dwarf is associated with different, possibly magnetic, phenomena. A number of studies \citep{2005ApJ...627..960B,Hallinan:2006vn,Hallinan:2008lr,2009ApJ...695..310B,McLean:2011qy} have suggested that some ultracool dwarfs have stable magnetic large-scale topologies, like dipoles or quadrupoles.  This could result in a large high-latitude feature. \citet{Hussain:2002qy} reviewed starspot lifetimes: While starspots on single, young main-sequence stars have typical lifetimes of a month, other types of starspots can have lifetimes of years or decades.  Furthermore, an active longitude may remain more heavily spotted for years, even as individual spots come and go on shorter timescales.  Of course, as \citet{Lane:2007yq} remark, a cooler starspot on an L dwarf might have different cloud properties than the normal photosphere.

\section{Flare Analysis\label{sec-flare}}

\subsection{Spectroscopy and photometry of the 29 July 2012 flares\label{geminiflare}}

\begin{figure}
\plotone{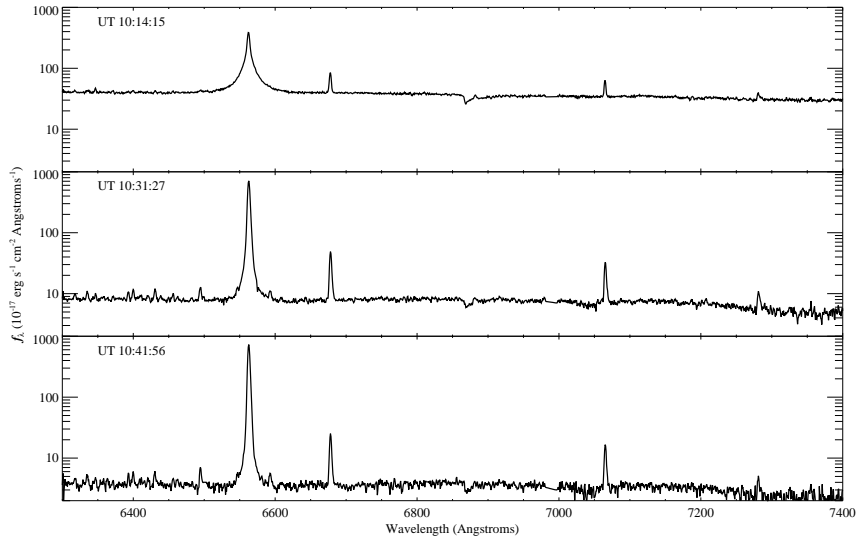}
\caption{The first three observed spectra during the main flare on UT Date 29 July 2012, with the L dwarf photosphere (Figure~\ref{fig-spectrum1}) subtracted off. Note the strong white light continuum, the broad H $\alpha$ that narrows, and the helium emission lines. The first spectrum has been adjusted upwards by a factor of 1.67 based on the timing of the flare in the Kepler photometry.  
\label{fig-flarespectra}}
\end{figure}

\begin{figure}
\epsscale{0.7}
\plotone{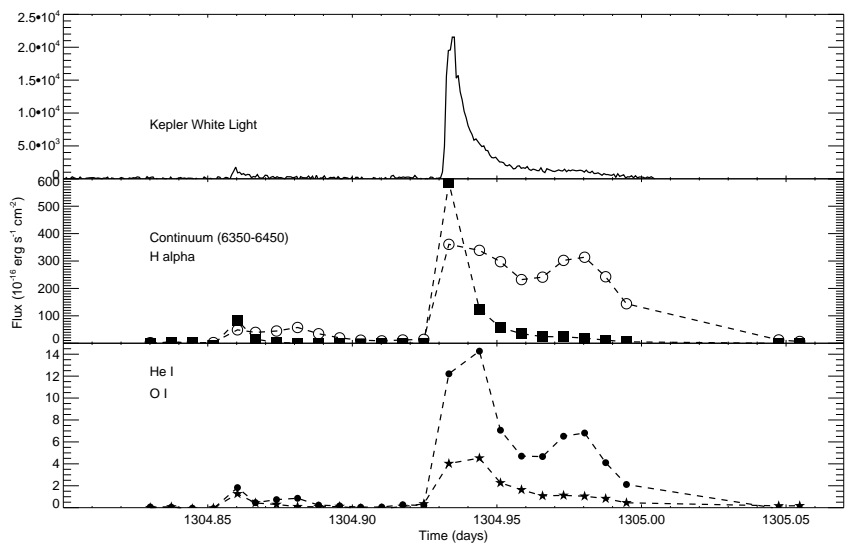}
\caption{Observations of the UT Date 29 July 2012 flares. Kepler white light and Gemini emission line flux as a function of time. The early, main, and later flares are evident. The line fluxes measured with Gemini at the start of the two flares (1304.86 and 1304.93) have been adjusted upwards based on the timing of the flare start times in the Kepler photometry.  The Kepler white light flux is calculated for the wavelength range 4370-8360 \AA; the ultraviolet plus optical flux is probably two to three times higher (Table~\ref{tab-calib}). The helium line at $6678$\AA~and the oxygen blend at $7774$\AA~are plotted. The continuum flux observed by in the Gemini spectra in the range 6350-6450\AA~after subtracting the quiescent spectrum from earlier in the night is also shown; its time dependence is very similar to the Kepler photometry.  
\label{fig-flaretime}}
\end{figure}

After the first forty minutes of Gemini data on 29 July 2012 show relatively quiescent H$\alpha$ emission, the subsequent spectra are dominated by a series of flares characterized by strong emission lines and a blue continuum. The first flare is in the ten minute exposure beginning at UT 08:29:28, followed by a more powerful ``main" flare at UT 10:14:15 (shown in Figure~\ref{fig-spectrum1}. Both of these flares are detected in the Kepler short cadence photometry: the first flare peaks at mission time 1304.8597 and the main flare peaks at mission time 1304.9353.  The first three main flare spectra, with the quiescent L dwarf photosphere subtracted, are shown in Figure~\ref{fig-flarespectra} showing decay of the blue continuum. The atomic emission lines as a function of time throughout the night are shown in Figure~\ref{fig-flaretime} along with the simultaneous Kepler short cadence photometry.\footnote{The timing gaps evident in Figure~\ref{fig-flaretime} should be explained. The Gemini observations were originally queue scheduled as three blocks of three hours each, with two blocks on 29 July 2012. The main flare begins in the 15th spectrum, the last of the first block. There is then a five minute gap before the next block of spectra begins. After eight of these spectra,  the telescope was diverted to a target of opportunity for another program. After slightly more than one hour, the queue returned to our program to take two more spectra. The lost observing time was made up on 8 August. The Kepler light curve stops due to a monthly data download.}  The Kepler light curve is in good agreement with the appearance of the spectroscopic white light continuum. Even with multi wavelength data at high time resolution (see \citealp{Gershberg:2005lr}), flare spectra have proven to be challenging to interpret.  Most notably, this L1 dwarf is able to produce a white light flare that is similar to that observed in M dwarf flare stars.  Indeed, the flare spectrum is very similar to \citet{Fuhrmeister:2008lr}'s flare in the M6 dwarf \object{CN Leo} (Wolf 359), including the white light, broad H$\alpha$, and atomic emission lines. The flare also resembles the most powerful observed flares in M7 \citep{2007AJ....133.2258S} and M9 \citep{1999ApJ...519..345L} dwarfs, demonstrating continuity in the flare properties across the M/L transition. The white light and emission line light curves are qualitatively very similar to the 28 March 1984 AD Leo flare described by \citet{Houdebine:1992fk} as ``impulsive/gradual."
 
The main (1304.9395) flare shows both atomic emission lines and continuum emission with a blue slope. The higher time resolution photometry shows that the flare began four minutes into the Gemini UT 10:14:15 exposure, and reached its Kepler maximum in the final two minutes. The ``white light" flare continuum spectrum is consistent with an $8000\pm2000$K blackbody. In \citet{Allred:2006qy}'s radiative hydrodynamic simulations, an electron beam injected at the top of a magnetic loop heats the chromosphere, producing emission lines in general agreement with observed flares, but does not penetrate to the denser photosphere, and white light emission is not produced. As \citet{Kowalski:2010gf} discuss in the case of a flare on the M4.5 dwarf \object{YZ CMi}, evidently the W1906+40 photosphere must somehow be heated to produce $\sim 8000$K continuum emission. This is also seen in the older models of \citet{Cram:1982mz}, where white light continuum emission requires moving the temperature minimum deeper and heating the photosphere (see their Models 4-6).  Free-free emission from a hot ($T \approx 10^7$ K) plasma is also consistent with our white light flare spectrum but is disfavored by \citet{Hawley:1992fr}'s analysis of M dwarf flares. The Kepler photometry and our calibration (Section~\ref{sec-kepflare}, Table~\ref{tab-calib}) indicates that the total (included extrapolated to ultraviolet) flare energy release during the Gemini 10:14:15UT exposure was $\sim 5 \times 10^{31}$ erg. The photometric light curve is typical of those observed in UV Ceti (M dwarf) flare stars as summarized by \citet{Gershberg:2005lr}: It is asymmetric, with a rapid rise, an initial fast decrease, and a subsequent slow decay which began when the flare was at $\sim 0.2$ of the maximum. The flare lasts 106 minutes in the Kepler photometry. Using the calibration discussed in detail in Section~\ref{sec-kepflare}, the observed flare energy through the Kepler filter during this time is $6.4 \times 10^{31}$ ergs, which may be extrapolated to $\sim 1.6 \times 10^{32}$ erg for the full ultraviolet/optical range. 

The white light in the main flare is accompanied by strong, very broad H$\alpha$ emission, especially evident in the top spectrum of Figure~\ref{fig-flarespectra}. Most stellar flares do not show much H$\alpha$ broadening \citep{Houdebine:1992fk}.
If modeled with two gaussians, as sometimes done for M dwarf flares \citep{Eason:1992fr,Fuhrmeister:2008lr}, the narrower gaussian component has Full Width at Half Maximum (FWHM) $\sim 6$ \AA~ and the broader gaussian component has FWHM $\sim 25$ \AA. Very broad H$\beta$ and bluer Balmer flare lines are usually attributed to Stark broadening, but \citet{Eason:1992fr} argued in the case of a \object{UV Ceti} flare that broad H$\alpha$ is better explained by turbulent motions. \citet{Fuhrmeister:2008lr} also interpret broad H$\alpha$ flare emission in the M6 dwarf \object{CN Leo} as due to turbulent motions, in addition to Stark broadening, possibly due to ``a chromospheric prominence that is lifted during the flare onset and then raining down during the decay phase." (Their observations have higher time and spectral resolution.) This interpretation could also be applied to the W1906+40 flare. \citet{Zirin:1973fj} noted broad (FWHM 12 \AA) H$\alpha$ emission from the kernels of a great solar flare; these are also responsible for the white light emission and show turbulent motions \citep{Zirin:1988lr}.
Detailed modeling should be pursued, however, as very broad H$\alpha$ emission wings are caused by Stark broadening in some \citet{Cram:1982mz} models along with the white light (Models 5, 6), while \citet{Kowalski:2011mz,Kowalski:2012fr} found broad H$\alpha$ {\it absorption} associated with a white light flare on \object{YZ CMi} reminiscent of an A-star. 

As the blue continuum and broad H$\alpha$ decays, the H$\alpha$ line stays strong but narrows. Atomic emission lines from \ion{He}{1} (6678\AA, 7065\AA, 7281\AA) and \ion{O}{1}  (7774\AA) are easily visible. These emission lines remain strong for two hours (Figure~\ref{fig-flaretime}). One hour after the main flare at mission times 1304.97-1304.99, H$\alpha$ strengthens again, perhaps due to another flare, creating a double peaked time series profile, though only a small increase in the spectroscopic blue continuum and Kepler photometry occur. Subtracting the quiescent photospheric spectrum, we also identify emission lines from neutral K and Na which fill in the core of the photospheric absorption features. These spectra are similar to red M dwarf flare spectra \citep{Fuhrmeister:2008lr}, and are understood as the result of the increase in electron density and heating of the chromosphere \citep{Fuhrmeister:2010fk}. The much longer duration of the emission lines is not surprising, and is consistent with solar flares \citep{Zirin:1988lr} and the statistics of SDSS spectra of M dwarf flares \citep{Kruse:2010uq,Hilton:2010ly}. The flare luminosity for the first hundred minutes in units of $10^{28}$ ergs is 480, 13, and 3.5 in H$\alpha$, \ion{He}{1} (6678 \AA), and \ion{O}{1} (7774\AA). During the time period covered by the first three Gemini spectra (Figure~\ref{fig-flarespectra}, 80\% of the Kepler luminosity was emitted but only 60\% of the \ion{O}{1}, 50\% of the \ion{He}{1} and 40\% of the H$\alpha$ luminosities.  

The first (1304.8597) flare was also a powerful one in its own right. In the first Gemini spectrum of the first flare, H$\alpha$ strengthens to equivalent width 30\AA, which we can correct to 42\AA~ given the timing of the Kepler photometry. In the next ten minute exposure, H$\alpha$ has increased to 55\AA. When the quiescent photosphere is subtracted, a blue continuum is present in the first exposure and the flare is clearly detected in the Kepler photometry. We are unable to say if this flare, which would otherwise be the strongest spectroscopically observed L dwarf flare, is a physically related to the main flare almost two hours (a fifth of a rotation period) later. Overall, the complex time evolution, with the first flare two hours before and a secondary flare an hour after the main flare, suggests the emergence and reconfiguration of a complex, multi-loop magnetic structure. This may be consistent with the L dwarf flare theory sketched out by \citet{Mohanty:2002lr} in which flux tubes generated in the warmer interior rise rapidly into the cooler atmosphere, but in any case the flares appear very similar to flares on hotter stars, suggesting similar mechanisms. Although probable flares in early L dwarfs have been noted before \citep{2003AJ....125..343L,2007AJ....133.2258S,2008ApJ...684.1390R}, they were seen in the H$\alpha$ line only. The 29 July flares are both stronger in H$\alpha$ than in those flares, in addition to showing other atomic emission lines and white light. 

\subsection{Flare statistics from Kepler photometry\label{sec-kepflare}}

\begin{figure}
\epsscale{0.7}
\plotone{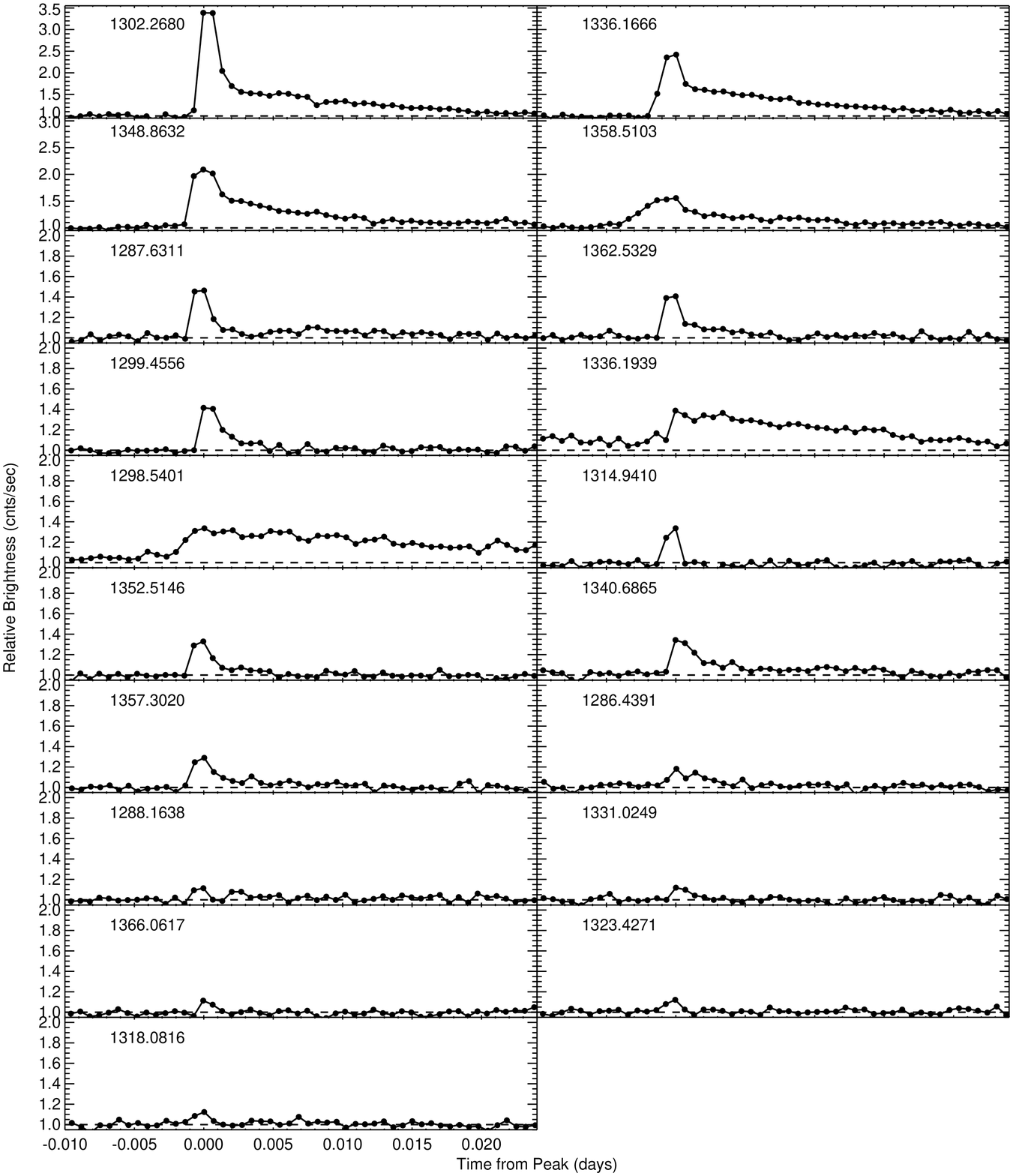}
\caption{Quarter 14 Kepler short cadence photometry of white light flares. Note the different scales for the upper plots. 
\label{fig-flarephotometry14}}
\end{figure}

There are a total of 21 candidate white light flares in the Quarter 14 short cadence data that peak at least 10\% above the average W1906+40 count rate and are elevated for two or more measurements (Table~\ref{tab-flares14}). Kepler time series photometry of these flares are shown in Figure~\ref{fig-flarephotometry14} The strongest candidate flares include data points flagged as likely cosmic rays by the Kepler pipeline, but we believe this is incorrect. The sharp rise of the two 29 July flares are also flagged as possible cosmic rays (see the discussion in \citealt{Jenkins:2010fk}), but the Gemini spectroscopy proves they are flares. The centroid of the main flare peak shifts by 0.4 pixels not because it is a off-center cosmic ray, but because the positions are wavelength dependent, and (blue) flare photons outnumber (red) photospheric photons more than three to one. As the flare decays the centroid returns to normal. We therefore accept flare-like light curves even if the sharp rise triggered the cosmic ray flag. 

\begin{figure}
\plotone{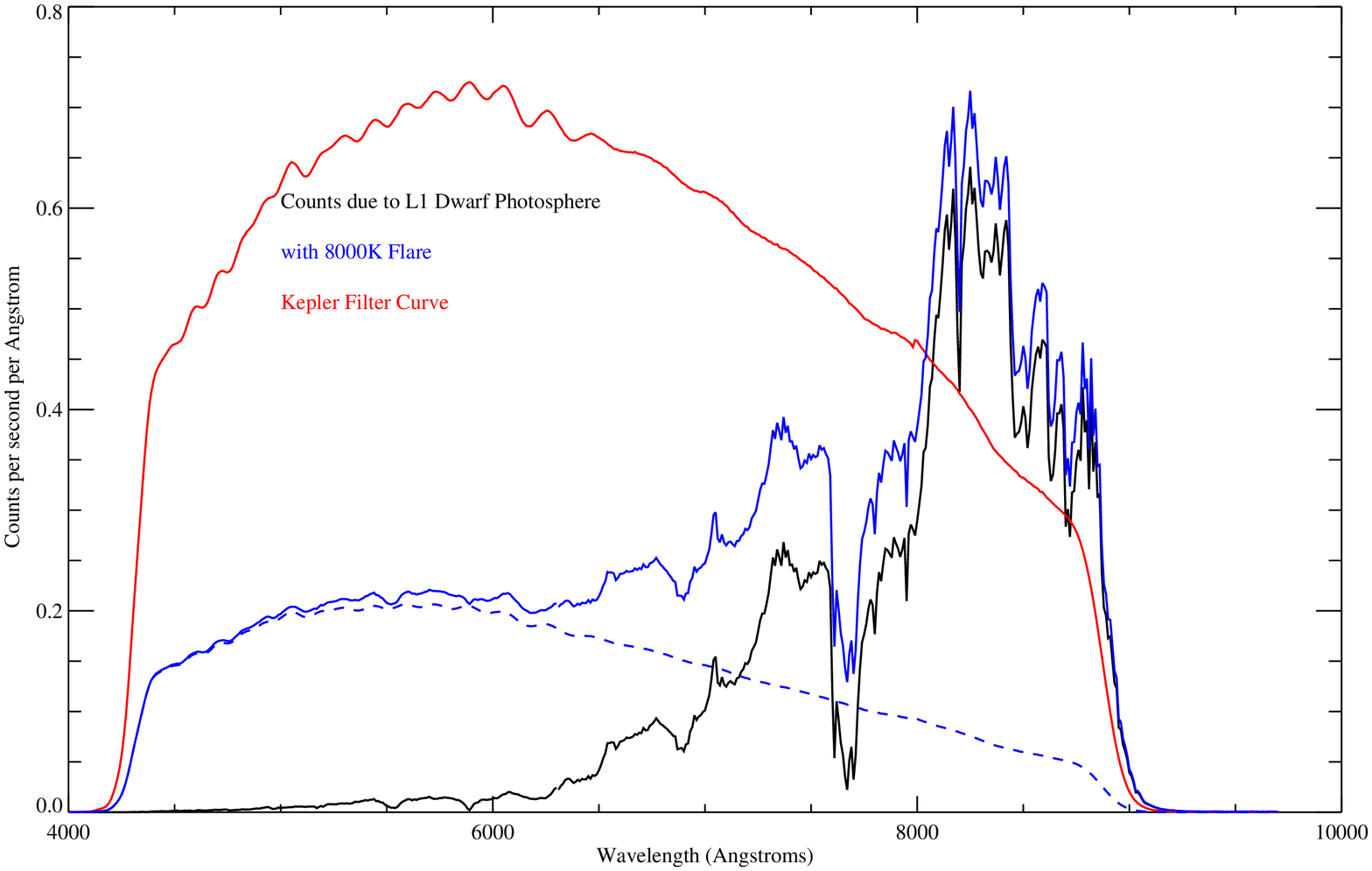}
\caption{Predicted count rates as a function of wavelength for the Kepler photometer. We use the L1 standard  2MASS J14392836+1929149 \citep{1999ApJ...519..802K}
for $\lambda > 6300$ \AA and the L0.5 dwarf 2MASS J07464256+2000321 \citep{2000AJ....119..369R} for $\lambda \le 6300$\AA~in order to completely cover the Kepler filter sensitivity range.
The counts are weighted by the average Kepler filter response \citep{2010ApJ...713L..79K}, which is also plotted by itself. The integrated count rate is normalized to give 660 electrons per second, which matches the pipeline reported count rate for Quarter 10. Also shown is a hypothetical 8,000K blackbody flare \citep{2011AJ....141...50W} that contributes an equal number of counts. \label{fig-filter} }
\end{figure}

It can be useful to discuss flares in terms of energies instead of relative counts.  \citet{2011AJ....141...50W} and \citet{Maehara:2012fr}  have estimated the energy of white light flares on G, K and M dwarfs observed by Kepler by approximating the flare as a 10,000K blackbody. The Kepler Mission filter is similar to $g$,$r$, and $i$ combined \citep{2010ApJ...713L..79K}: As illustrated in Figure~\ref{fig-filter}, the L dwarf photosphere contributes mainly at the very reddest wavelengths, but hot flares will contribute through the entire filter. Indeed, the effective wavelength for the Kepler photometry for W1906+40 is 7990\AA~rather than $\sim 6200$\AA~for hot sources. In Table~\ref{tab-calib}, we present energy calculations of hypothetical flares that have the same observed count rate as W1906+40, assuming the flares radiate isotropically. We use the $i$ magnitude, the distance derived in Section~\ref{optical}, and the L1 dwarf spectrum shown in Figure~\ref{fig-filter} to estimate that the L dwarf luminosity is $1.4 \times 10^{28}$ erg s$^{-1}$  in the wavelength range 4370\AA~to 8360\AA. The blackbody energies for equal count rates, $2.1 \times 10^{28}$ erg s$^{-1}$, are 50\% higher because the average energy per photon is larger. Thus, to double the observed Kepler count rate, one needs a flare of $1.3 \times 10^{30}$ erg (over that wavelength range) during a short cadence exposure and $3.8 \times 10^{31}$ erg during a long cadence exposure. Most of the flare energy will be emitted at shorter optical and ultraviolet wavelengths. The corresponding bolometric luminosities for the blackbodies are 4.4 to $6.4 \times 10^{28}$ erg s$^{-1}$.  Real flares are not blackbodies, so we also estimated the total luminosities by consulting Table 6 of \citet{ Hawley:1991uq}, which gives energies from 1200\AA to 8045\AA~ for the impulsive and gradual phases of great flare of 12 April 1984 on \object{AD Leo}. The results are close to the blackbody approximations.   

Seven of the flares last at least an hour. Most rise to their peak value within a few minutes, and drop to half their peak flux in 1-3 minutes. We list the rise time to the peak in Table~\ref{tab-flares14}) as $t_r$, the time to drop to half the flux as $t_{1/2}$, and the total length of the flare as $t_L$.  The flare at mission day 1298.5401 is an exception with, a 23 minute rise time and slow decay. Overall, both the typical shape and the variety of the light curves is consistent with stellar flares \citep{Gershberg:2005lr}.  We also list the peak long cadence count rate in Table~\ref{tab-flares14}; most short cadence flares are enhancements of 2\% or less in the 30 minute data and therefore would not be significant detections without the short-cadence data. The cosmic ray flagging may have a significant effect on real flares in the long cadence data: The long cadence flux for the Gemini-confirmed 1304.9353 flare is half that we obtain by integrating the short cadence flux, and the discrepancy is due to the removal of ``cosmic rays" in the long cadence pipeline. We have searched the Quarters 10-13 data for other candidate flares; however, flares are easily confused with cosmic rays, and we have concluded that we cannot reliably distinguish between flares and cosmic rays using the long cadence data. 
We nevertheless mention two interesting candidates. At mission time 1148.407, the target increases by 48\%, decaying to 10\% in the next exposure; this would represent a flare of energy $\sim 1.4 \times 10^{32}$ ergs. (This is the time period also plotted in Figure~\ref{fig-radioregion}.) These points are not flagged as cosmic rays and the pixel data appears consistent with the stellar PSF, so it is likely to be a real flare. The most powerful long-cadence--only flare candidate is at mission time 1039.049: W1906+40 nearly doubles in brightness, then decays over the next 2 hours back to the normal brightness. The pipeline, however, flags the initial increase as a possible cosmic ray strike, and if this is correct, the following ``decay" simply represents the return of the CCD's sensitivity to normal. Flare rates should should be viewed with caution for the lowest energy flares, since the contribution of a long-lived tail can be lost in the noise, and flares with low peaks will be missed altogether even in the short cadence data. There is no apparent dependence of the flares on the rotation phase. It is also notable that four of the eight most powerful Quarter 14 flares might be related, the pair 1304.8597/1304.9353 and the pair 1336.1666/1336.1939.
  
\begin{figure}
\epsscale{0.8}
\plotone{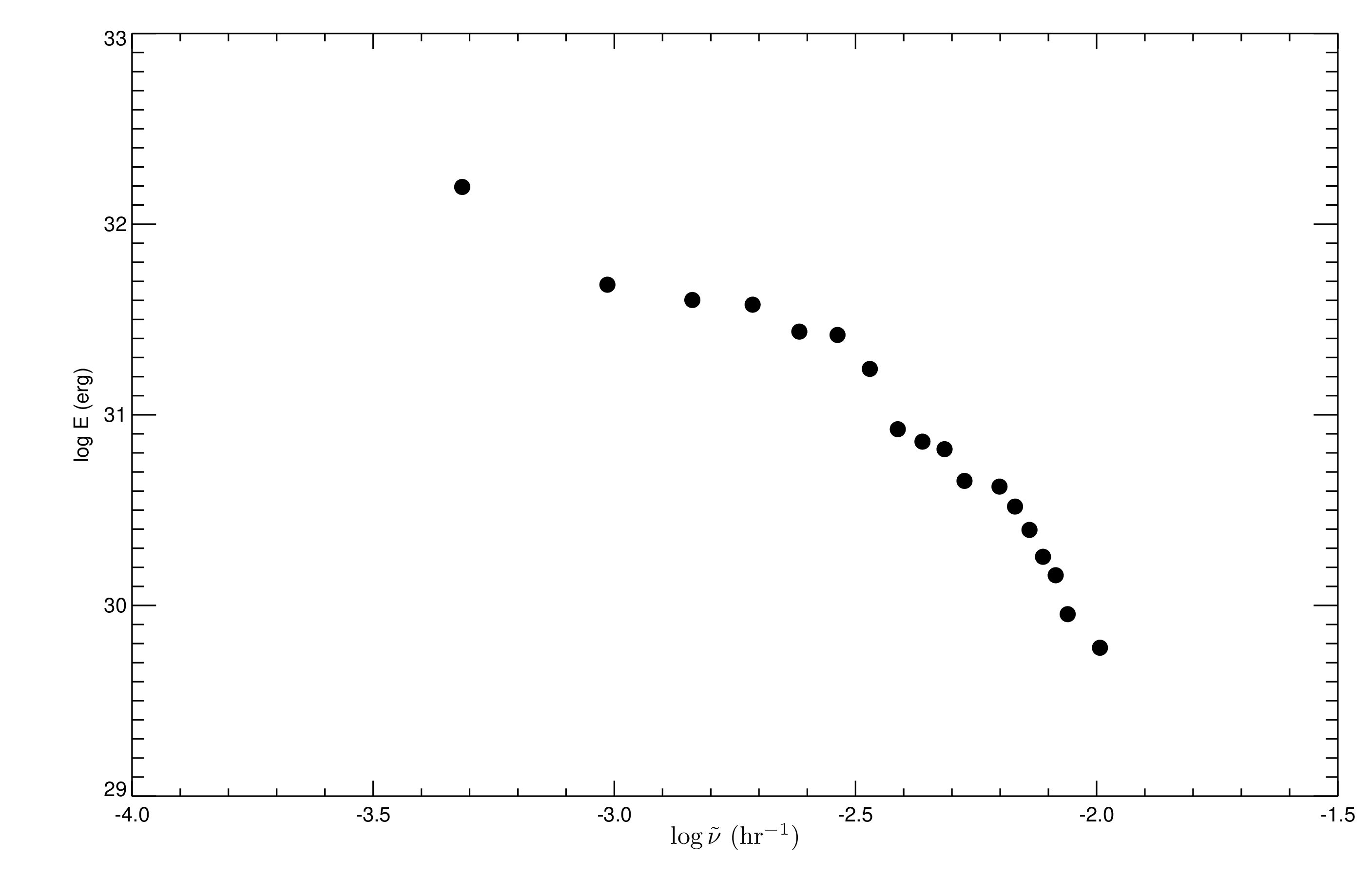}
\caption{Energy spectra of white light flares on W1906+40 for short cadence Quarter 14 data (solid circles). The flare energies have been calculated for the 8000K blackbody (see Table~\ref{tab-calib}); the observed energies are 2.4 times smaller.  The most energetic flare is the one observed spectroscopically (Figures~\ref{fig-flarespectra} and~\ref{fig-flaretime}). 
\label{fig-flarefrequency}} 
\end{figure}

In Figure~\ref{fig-flarefrequency}, we plot the flare energy spectrum \citep{Lacy:1976qy} as the estimated flare energy as a function of the cumulative frequency ($\tilde{\nu}$). Additional monitoring would be needed to improve the frequency estimates and to check if the flare rate varies. Comparison to well studied dMe flare stars \citep{Lacy:1976qy,Gershberg:2005lr} indicates the flare energy spectrum is similar to that observed in flare stars but shifted downwards for W1906+40: Flares are $\sim 10-100$ times less energetic at a given frequency, or alternatively flares of a given energy are less frequent on W1906+40 than on dMe flare stars.  On the other hand, the white light flare rate on W1906+40 is comparable to that of the Sun \citep{Neidig:1983kx,Gershberg:2005lr}, despite the L dwarf's much lower effective temperature and surface area.  The observed rate for flares with energy $>10^{31}$ ergs is $10^{-3}$ to $10^{-2.5}$ hr$^{-1}$, or 1-2 per month.  For comparison, the \citet{2010ApJ...709..332B} limit on H$\alpha$-only L dwarf flares is a cumulative frequency of $\log \tilde{\nu} (\rm{(hr}^{-1}) \lesssim -1.4$; those flares are too weak to be detected in the broad Kepler filter, because a change of 10\AA~would change the photometry by only $\sim0.0015$\%.  

\section{Conclusion}

Our initial monitoring of W1906+40 has revealed long-lived surface features and a well-defined periodicity due to rotation. Despite its cool photosphere, this L1 dwarf maintains an active and variable chromosphere, emits in the radio, and produces powerful flares that rival white light flares in the Sun. A major limiting factor in the analysis of the periodic component is our lack of knowledge of the spot-to-star flux ratio. During our Guest Observer Cycle 4 program, we are monitoring W1906+40 with a number of ground-based telescopes to search for changes correlated with Kepler data and to determine if the surface features change on multi-year timescales. These observations may be able to reveal the nature of the surface features, and whether they are hotter or cooler than the surrounding photosphere. 

\acknowledgments

We are grateful to Martin Still for approving the Kepler observations and to the USNO infrared parallax team for allowing us to use the preliminary parallax measurement. We thank Gibor Basri and Peter Plavchan for helpful discussions of Kepler photometry, Kamen Todorov and David Hogg for discussions of MCMC methods, Mark Giampapa, Suzanne Hawley, Adam Kowalski, Dermott Mullan, and Lucianne Walkowicz for discussions of stellar flares, and the anonymous referee for helpful comments.  

This paper includes data collected by the Kepler mission. Funding for the Kepler mission is provided by the NASA Science Mission directorate. 
The material is based upon work supported by NASA under award No. NNX13AC18G.  EB acknowledges support from the National Science Foundation through Grant AST-1008361. The National Radio Astronomy Observatory is a facility of the National Science Foundation operated under cooperative agreement by Associated Universities, Inc. Based on observations obtained at the Gemini Observatory, which is operated by the Association of Universities for Research in Astronomy, Inc., under a cooperative agreement with the NSF on behalf of the Gemini partnership: the National Science Foundation (United States), the Science and Technology Facilities Council (United Kingdom), the National Research Council (Canada), CONICYT (Chile), the Australian Research Council (Australia), Minist\'{e}rio da Ci\^{e}ncia, Tecnologia e Inova\c{c}\~{a}o (Brazil) and Ministerio de Ciencia, Tecnolog\'{i}a e Innovaci\'{o}n Productiva (Argentina) Some of the data presented herein were obtained at the W.M. Keck Observatory, which is operated as a scientific partnership among the California Institute of Technology, the University of California and the National Aeronautics and Space Administration. The Observatory was made possible by the generous financial support of the W.M. Keck Foundation.  This publication makes use of data products from the Wide-field Infrared Survey Explorer, which is a joint project of the University of California, Los Angeles, and the Jet Propulsion Laboratory/California Institute of Technology, funded by the National Aeronautics and Space Administration. Some of the data presented in this paper were obtained from the Mikulski Archive for Space Telescopes (MAST). STScI is operated by the Association of Universities for Research in Astronomy, Inc., under NASA contract NAS5-26555. Support for MAST for non-HST data is provided by the NASA Office of Space Science via grant NNX09AF08G and by other grants and contracts.

{\it Facilities:}  \facility{Gemini:Gillett}, \facility{Keck:II}, \facility{Kepler},  \facility{USNO:61in},  \facility{VLA}

\bibliographystyle{apj}

\begin{deluxetable}{lcc}
\tablewidth{0pc}
\tabletypesize{\footnotesize}
\tablenum{1} \label{tab1}
\tablecaption{WISEP J190648.47+401106.8}
\tablehead{
\colhead{Parameter} &
\colhead{W1906+4011} & \colhead {Units} }
\startdata
Sp. Type &L1 & optical \\
Sp. Type &L1 & near-infrared\tablenotemark{a} \\
$i$ & $17.419\pm 0.006$ & mag \tablenotemark{a} \\
$K_s$ & $11.77\pm 0.02$ & mag \tablenotemark{a} \\
$\pi_{abs}$ & $61.15 \pm 1.30$ & mas \\
$\mu$ & $476.2 \pm 2.9$ &mas yr$^{-1}$ \\
$\theta$ & $111.28 \pm 0.17$  & degrees\\
$v_{tan}$ & $36.9 \pm 0.8$ & km s$^{-1}$ \\
$v_{rad}$  & $-22.6 \pm 0.4$ & km s$^{-1}$  \\
U & -5.5 & km s$^{-1}$  \tablenotemark{b} \\
V  & -11.6 & km s$^{-1}$ \\
W  & -41.3 & km s$^{-1}$ \\
$v \sin i$  & $11.2 \pm 2.2$ & km s$^{-1}$ \\
Period & 0.37015 & days\\
Amplitude & 1.5\%&  \\
$\log L_{bol}/L_\odot $ &  $-3.67 \pm 0.03$ & \\
$R$  &$ 0.92 \pm 0.07$ & $R_J$ \\
$\sin i$ & $>0.59$ & \\
Radio $\nu L_\nu$ & $(4.5  \pm 0.9) \times 10^{22}$ & erg s$^{-1}$ \\
Radio $\alpha$ & $ -2.0 \pm 1.4$ &  \\
$\langle$H$\alpha$ EW$\rangle$  & $4$ & \AA  \\
$\langle F_{\rm{H}\alpha} \rangle$  & $3 \times 10^{-16}$ & erg s$^{-1}$cm$^{-2}$  \\
$\log {\langle L_{\rm{H}\alpha} \rangle/L_{bol}}$  & -5.0 &   \\
$\log {\nu L_\nu}/L_{bol}$  & -7.3 &   \\
\enddata
\tablenotetext{a}{From \citet{Gizis:2011lr}}
\tablenotetext{b}{U is positive towards Galactic center}
\end{deluxetable}

\begin{deluxetable}{lcccc}
\tablewidth{0pc}
\tabletypesize{\footnotesize}
\tablenum{2} \label{tab-spot}
\tablecaption{Sample Single Circular Dark Spot Models}
\tablehead{
\colhead{Parameter} & 
\colhead{$i = 50^\circ$} &
\colhead{$i = 60^\circ$} &
\colhead{$i = 70^\circ$} &
\colhead{$i = 80^\circ$} }
\startdata
$\sin i$ & 0.77 & 0.87 & 0.94 & 0.98 \\
Median Observed Flux & 1.000 & 1.000 & 1.000 & 1.000\\
Unspotted Flux &1.012 & 1.012  & 1.011 & 1.011 \\
Latitude (deg) & 73 & 77 & 81 & 83 \\
Spot Radius (deg) & 8.0 & 9.9 & 13 & 20 \\
\enddata
\end{deluxetable}

\begin{deluxetable}{lcccc}
\tablewidth{0pc}
\tabletypesize{\footnotesize}
\tablenum{3} \label{tab-calib}
\tablecaption{W1906+40 Kepler Calibration}
\tablehead{
\colhead{Spectrum} &  
\colhead{Count Rate} &
\colhead{$L$ ({4370-8360 \AA})} &
\colhead{$L_{bol}$} \\
\colhead{} & 
\colhead{cnts s$^{-1}$} &
\colhead{$10^{28}$ erg s$^{-1}$} &
\colhead{$10^{28}$ erg s$^{-1}$} 
} 
\startdata
L1 photosphere (W1906+40) & 1  & 1.4 & 82 \\
Flare (6000K Blackbody) & 1 & 2.0 & 4.4 \\
Flare (8000K Blackbody) & 1 & 2.1 & 5.0 \\
Flare (10000K Blackbody) & 1 & 2.1 & 6.4 \\
Flare (Scaled AD Leo Impulsive)\tablenotemark{a} & 1 & 2.1 & 6.8 \\
Flare (Scaled AD Leo Gradual)\tablenotemark{a}  & 1  & 2.1 & 5.2 \\
\enddata
\tablenotetext{a}{Based on \citet{Hawley:1991uq}. See text}
\end{deluxetable}

\begin{deluxetable}{rrcrrrll}
\tablewidth{0pc}
\tabletypesize{\footnotesize}
\tablenum{4} \label{tab-flares14}
\tablecaption{Flares Observed in Quarter 14}
\tablehead{
\colhead{Mission Time} & \colhead{Peak} & \colhead{$E_{tot}$} & \colhead{$t_L$}& \colhead{$t_{r}$ } & \colhead{$t_{1/2}$ } & Peak$_{LC}$ & \colhead{Notes} \\
\colhead{days} & \colhead{cnts s$^{-1}$} & \colhead{$10^{30}$ erg} & \colhead {min} & \colhead {min} & \colhead {min} & \colhead{cnts s$^{-1}$} &  \colhead{}  }
\startdata
 1304.9353 &   4.27 &  160~ & 106 & 7 & 5 & 1.35  & Gemini spectra\\ 
 1298.5401 &   1.34 &  48~ & 160 & 23  & 22 & 1.16  \\ 
 1336.1666 &   2.42 &  40~ & 37 & 3 & 2 & 1.17 & Ended by 1336.1939\\ 
 1348.8632 &   2.09 &   38~ & 75 & 5 & 3 & 1.10 \\ 
 1358.5103 &   1.56 &   27~ & 76 & 9 & 3 & 1.08 \\ 
 1336.1939 &   1.39 &   26~ & 64 & 4 & 17 & 1.08 \\ 
 1302.2680 &   3.39 &   17~ & 106 & 2 & 2 & 1.02 \\ 
 1304.8597 &   1.27 &   8.4~ & 60 & 3 & 1 & 1.04 & Gemini spectra, precedes 1304.9353\\ 
 1287.6311 &   1.46 &   7.2~ & 26 & 2 & 1 & 1.04  & \\ 
 1340.6865 &   1.34 &   6.6~ & 18 & 1 & 3 & 1.02  \\ 
 1362.5329 &   1.41 &   4.5~ & 12 & 2 & 1 & 1.01 \\ 
 1357.3020 &   1.29 &   4.2~ & 19 & 2 & 1 & 1.00  \\ 
 1299.4556 &   1.42 &   4.2 & 8 & 1 & 2 & 1.01 \\ 
 1352.5146 &   1.33 &    3.3 & 9 & 2 & 1 & 1.00 \\ 
 1286.4391 &   1.18 &    2.5 & 12 & 2 & 3 & 1.00 \\ 
 1314.9410 &   1.34 &    1.8 & 2 & 2 & 1 & 1.00 \\ 
 1288.1638 &   1.11 &    1.4 & 10 & 2 & 1 & 1.01 \\ 
 1331.0249 &   1.12 &    0.9 & 4 & 1 & 2 & 1.00 \\ 
 1366.0617 &   1.11 &    0.6 & 2 & 1 & 1 & 1.00 \\ 
 1323.4271 &   1.12 &    0.6 & 3 & 2 & 1 & 1.00 \\ 
 1318.0816 &   1.12 &    0.6 & 3 & 2 & 1 & 1.00 \\ 
\enddata
\end{deluxetable}

\end{document}